\documentclass[
  a4paper,twocolumn,floatfix,rmp,showkeys,superscriptaddress
]{revtex4}

\usepackage{graphicx}
\usepackage{amsmath}
\usepackage{amsfonts}

\bibstyle{revtex}
\bibpunct{(}{)}{,}{}{}

\begin{document}
\bibliographystyle{agsm}

\title{
        Minimal model of self-replicating nanocells: 
        a physically embodied information-free scenario
}

\providecommand{\ICREA}{
        ICREA-Complex Systems Lab, Universitat Pompeu Fabra (GRIB), 
        Dr Aiguader 80, 08003 Barcelona, Spain
}
\providecommand{\SFI}{
        Santa Fe Institute, 1399 Hyde Park Road, Santa Fe NM 87501, USA
}

\author{Harold Fellermann}
\email{harold.fellermann@upf.edu}
\affiliation{\ICREA}
\author{Ricard V. Sol\'e}
\email{ricard.sole@upf.edu}
\affiliation{\ICREA}
\affiliation{\SFI}

\begin{abstract}
The building of minimal self-reproducing systems with a physical 
embodiment (generically called protocells) is a great challenge, 
with implications for both theory and applied sciences. Although the 
classical view of a living protocell assumes that it includes 
information-carrying molecules as an essential ingredient, a 
dividing cell-like structure can be built from a 
metabolism-container coupled system, only. An example of such a 
system, modeled with dissipative particle dynamics, is presented 
here. This article demonstrates how a simple coupling between a 
precursor molecule and surfactant molecules forming micelles can 
experience a growth-division cycle in a predictable manner,
and analyzes the influence of crucial parameters on this 
replication cycle. Implications of these results for origins of 
cellular life and living technology are outlined.
\end{abstract}


\maketitle

\section{Introduction}

The transition from non-living to living systems covers a broad
spectrum of increasingly complex organization \cite{Smi:1995}. 
One of such first steps in this transition leads from ordinary 
chemical auto-catalysis to self-replication. The latter can be
distinguished from the former by the existence of self-bounded
entities which produce copies of themselves, rather than a mere 
increase in chemical concentration. Therefore, self-replication 
relies on organization principles unlikely to be found in 
homogeneous solutions. Within the last years increasing attention 
has been payed to the possibility of building small-scale 
protocells, in particular using a bottom-up approach \cite{Szo:2001}
where the building blocks (not necessarily from biotic 
origin) would assemble spontaneously and, under appropriate 
conditions, develop a growth-fission cycle. While extensive research has been performed 
on the self-reproducing capabilities of bio-polymers
\cite{Tji:1990,Kie:1986}, even much simpler systems can be driven 
into dynamics that we identify as self-replication. In this context, 
micelles have been proposed to serve as life-like structures able 
to undergo self-reproduction \cite{Bac:1992}. 

These micellar systems can be considered to be simpler in 
organization than bio-polymers because they lack any genetic 
information that could be passed from one generation to the next.
In this context, protocellular entities lacking information would 
be under the umbrella of Oparin's views of life origins  \cite{Opa:1936}, 
who suggested that primitive self-replicating vesicles would have 
predated information-based cells. Similarly, other authors have 
advocated for this scenario under a computational perspective. Under 
this view, Dyson \cite{Dys:1999} indicated that current cells involved 
both software (DNA and RNA) and hardware (protein machinery). Although 
hardware can exist under the absence of software, the opposite is not 
allowed to occur. We can add to this picture of cells that the presence 
of a container is a very important piece for the hardware to properly 
work: only when the pieces are close together we can expect the 
machinery to operate.

\begin{figure}
\includegraphics[width=7.5cm]{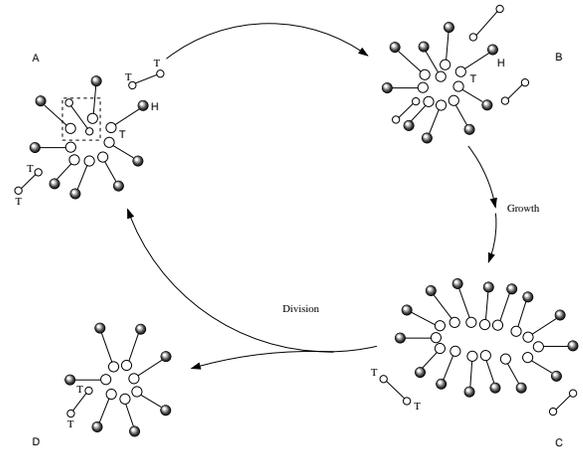}
\caption[Schematic of nanocell replication cycle]{
The basic model of nanocell replication explored in this 
paper. Here small-sized micelles are formed by amphiphiles (here 
indicated as $H-T$ connected pairs of balls). These amphiphiles have 
a hydrophilic head ($H$) and a hydrophobic tail ($T$). 
Precursor molecules are also shown as two connected, smaller open 
balls, both of them of hydrophobic character. Under the presence of  
catalyically active amphiphiles, precursors are transformed into
additional amphiphiles. Incorporation of many such building blocks
allows the nanocell to grow in size. When a critical size has been
reached, the nanocell looses its stability and divides into two smaller
aggregates thereby closing the replication cycle.
}
\label{fig_nanocell}
\end{figure}

The nanocellular system discussed here is based on a micelle that is 
coupled to a minimal metabolism     (figure \ref{fig_nanocell}). 
The system is constantly driven away from equilibrium by the
supply of precursors, which are supposed to have higher internal
energy than the surfactants that build up the micelles.
The metabolism transforms precursor molecules into new building 
blocks of the nanocell. The metabolic turnover is thereby enhanced 
by its own outcome or---in other words---the metabolism is an 
auto-catalytic turnover of precursors into new surfactants. This 
system resembles one studied experimentally by Bachmann et al. 
\cite{Bac:1992}. However, in their work the catalytic effect is a 
feature of the micelles (probably due to milieu effects), while, in 
our system, it is a feature of the molecules themselves.

Either way, the replication mechanism of micelles envisioned both 
in Bachmann's as well as our system is the following: micelles 
incorporate hydrophobic precursor molecules 
where they are afterwards transformed into new surfactants. Due to 
this process, the number of surfactants increases, while at the 
same time, the volume of the hydrophobic core becomes smaller. It 
is assumed, that when a critical ratio of surfactants versus core 
volume is passed, the aggregates become unstable and will divide 
into two daughter cells. Whether or not the experimental system 
follows this pathway has, to our knowledge, never be clarified.
However, once such a replication mechanism is considered the
basis for a more complicated protocell design, knowledge of the 
exact replication kinetics becomes crucial.

Together with experimental approaches aimed to finding the conditions 
for protocell replication to occur, there is an increasing need 
of appropriate, well-grounded theoretical and computational models. 
Such models can help understanding the constraints that might operate 
in the self-assembly of micelles and other molecules and how they 
can properly trigger growth and splitting. Our work is a first step 
in this direction.

\section{DPD nanocell model}

In this paper we consider a {\em dissipative particle dynamics} 
(DPD) approach to modeling embodied protocells employing a 
physically and chemically simplified formalism. This is one possible approach 
that can be used in order to tackle the complexities of molecular aggregates. 
Other approaches include Molecular Dynamics (MD), Brownian Dynamics (BD) 
and Monte Carlo algorithms. Each of these methods has its own advantages 
and drawbacks \cite{Hee:1990,Bin:1997} and a compared 
analysis has been presented elsewhere \cite{Che:2004}. 

Previous work done by Ono and Ikegami involved a lattice-based, protocell dynamics 
\cite{Ono:1999}. These structures display a special type of cell-like 
replication. While they are 
remarkable in their self-organizing behavior, the underlying rules 
of interaction and the special properties of the membrane-like 
structures restrict their relevance to the arena of artificial 
life. By using more appropriate molecular interactions within a 
three-dimensional, water-filled environment (see for example Rasmussen and Nilsson's  
cellular automata approach to self-assembly \cite{Ras:2003}) 
we seek to provide the (as far as we know) first computational evidence that such a 
simple protocellular cycle is feasible.

\subsection{Dissipative particle dynamics}

DPD is a coarse grained, particle based simulation technique
comparable to Brownian Dynamics. It was proposed by 
Hoogerbrugge and Koelmann \cite{Hoo:1992} and gained significant 
theoretical support in the late 1990s 
\cite{Esp:1995,Gro:1997,Mar:1998}. In the context of
biological systems, DPD models have been successfully used to 
capture the dynamics of membranes \cite{Ven:1999}, vesicles 
\cite{Yam:2002,Yam:2003} and micelles \cite{Gro:2000,Yua:2002}.

A DPD simulation consists of a set of $N$ particles which are 
described by their type, mass $m_i$, position $\mathbf r_i$, and 
momentum $\mathbf q_i = m_i\mathbf v_i$. These particles---usually 
called {\em beads} throughout the literature---are not meant to 
represent individual atoms. Instead, they represent groups of 
atoms within a molecule (like several $CH_2$ groups within a 
hydrocarbon chain) or even a group of small molecules such as water.

Newton's Law of motion is used to determine the trajectory of each 
individual bead:
\begin{equation}
        \frac{d^2\mathbf r_i}{dt^2} = \frac 1 m \mathbf F_i
        \label{eqn_newton}
\end{equation}
The force $\mathbf F_i$ that acts on particle $i$ is expressed as 
the superposition of pairwise interactions
\begin{equation}
        \mathbf F_{i} = \sum_{j=1}^N \mathbf F_{ij}
\end{equation}
In Newtonian dynamics, the (central) force $F^C_{ij}$ can be expressed as the negative 
gradient of a potential $\phi_{ij}$, namely 
\begin{equation}
F^C_{ij}=-\nabla\phi_{ij} 
\end{equation}
The resulting dynamics are conservative and obey the Hamiltonian 
\begin{equation}
H = \frac 1 2 \sum_{i=1}^N m_i \mathbf v_i 
  + \frac 1 2 \sum_{i,j=1}^N \phi_{ij}
\end{equation}
While this approach is undertaken in molecular dynamics 
simulations, coarse grained simulation techniques try to aggregate 
some of the molecular degrees of freedom by the use of the so called Langevin
formalism: additional forces  $\mathbf F^D_{ij}$ and 
$\mathbf F^R_{ij}$ are added to the conservative force to express
friction and thermal motion. They introduce energy flows between
the explicitly modeled mesoscale and an the underlying 
microscale. Together, they act as a thermostat to regulate the
effective temperature, i.~e. mean velocity, of the system.

The thermostat used in the DPD formalism is given by the equation
\begin{equation}
\mathbf F^D_{ij} + \mathbf F^R_{ij} = 
        \left(
                \eta \omega(r_{ij})(\mathbf n_{ij}\cdot\mathbf v_{ij}) +
                \sigma \omega^2(r_{ij})\xi_{ij}
        \right)\mathbf n_{ij}
        \label{eqn_thermostat}
\end{equation}
where $r_{ij}=\left|\mathbf r_i-\mathbf r_j\right|$ is the distance,
$\mathbf n_{ij}=(\mathbf r_i-\mathbf r_j)/r_{ij}$ the (unit)
direction, and $\mathbf v_{ij} = \mathbf v_i-\mathbf v_j$ the
relative velocity between beads $i$ and $j$. $\eta$ is the friction
coefficient of the fluid and $\sigma$ the amplitude of thermal
motion. $\xi_{ij}$ is a random variable with $\xi_{ij} = \xi_{ji}$,
and otherwise Gaussian statistics.
$\omega$ is a distance weighing function usually defined as
\begin{equation}
        \omega(r) = 2\left(1-\frac r {r_c}\right)
\end{equation}
where $r_c$ is a cutoff range used to limit the maximal range of 
all interactions for performance reasons. It can be shown that
the equilibrium temperature of the system tends towards
$k_bT =\sigma^2/2\eta$ \cite{Esp:1995}.

While there is a variety of other thermostats used in coarse grained
particle simulations, the above mechanism is unique in that it
both conserves linear and angular momenta and fulfills the 
fluctuation-dissipation theorem. As a consequence of the
former, the resulting dynamics are consistent with the 
Navier-Stokes equations and hence preserve laminar flow properties
of the system. The latter property ensures an energy
distribution in the system following Maxwell-Boltzmann statistics.
The overall dynamics, therefore, capture both hydrodynamic
and thermodynamic traits of the systems.

In almost all DPD studies, the conservative force is derived
from a {\em soft-core potential} of the shape
\begin{equation}
        \phi_{ij}(r) = 
        \left\{
        \begin{array}{cl}
                \frac 1 2 a_{ij}r_c \left(1-\frac r {r_c}\right)^2 
                & \mbox{ if } r \le r_c \\
                0 & \mbox{ if } r > r_c
        \end{array}
        \right.
\end{equation}
The potential energy expressed by $\phi_{ij}$ should not be
understood as the mechanical energy, i.~e. enthalpy, of the system,
but rather as its free energy contribution \cite{Pag:2001}. 
Following this rationale, the interaction parameters $a_{ij}$ are 
used to express dissimilarities of substances due to high 
enthalpy as well as entropy contributions, respectively. Therefore, 
they can be related to Flory-Huggins coefficients known from 
polymer theory.

For the study of lipids and surfactants, covalent bonds between
beads are commonly introduced as harmonic spring forces: on top of
the above interactions, bonded beads interact according to the
potential
\begin{equation}
        \phi_{ij}^B(r) = \frac {b r_b} 2 \left(1-\frac r {r_b}\right)^2
\end{equation}
where $b$ is the strength and $r_b$ the optimal distance of covalent
bonds. As usual, we use $r_c$, $m$, and $k_bT$
as units of space, mass, and energy, respectively. The time unit
follows from equation \ref{eqn_newton} as $\tau=\sqrt{m/k_bT}r_c$.

To model the system under consideration, we define beads of type 
$\mathbf W$ (water), $\mathbf H$ (hydrophilic ``heads'') and 
$\mathbf T$ (hydrophobic ``tails'' of amphiphiles) with interaction
parameters taken from \cite{Gro:2000}, unless otherwise specified:
\begin{center}
\begin{tabular}{r|ccc}
            & $\mathbf W$ & $\mathbf H$ & $\mathbf T$ \\
\hline
$\mathbf W$ &   $25k_bT$  &   $15k_bT$  &   $80k_bT$  \\
$\mathbf H$ &   $15k_bT$  &   $35k_bT$  &   $80k_bT$  \\
$\mathbf T$ &   $80k_bT$  &   $80k_bT$  &   $15k_bT$  
\end{tabular}
\end{center}
All beads have mass $1m$. Precursor molecules are modeled as dimers 
of bonded $\mathbf T$ beads, surfactants as dimers of one 
$\mathbf T$ and one $\mathbf H$ bead. Here we have: $b=125k_bT$, $r_b=0.5r_c$ for 
all covalent bonds. 

The objective behind this parameter set is to model surfactants 
that form spherical micelles. To achieve this, the effective head 
area must be large compared to the volume of the hydrophobic core 
(packing parameter $1/3$). This is expressed by 
$a_{\mathbf{TT}} < a_{\mathbf{WW}} < a_{\mathbf{HH}}$. Furthermore, 
surfactant heads have a high affinity to water 
($a_{\mathbf{HW}} < a_{\mathbf{WW}}$), which is usually due to 
charges in the hydrophilic groups of the molecules. This assumption 
ensures that aggregates with high surface area (spherical micelles) 
are prefered over aggregates with less surface area (rod-like micelles)
in the process of total energy minimization.

\subsection{Incorporation of chemistry}
The metabolic reaction under consideration takes the following form
\begin{equation}
        \mathbf{T-T} \longrightarrow \mathbf{H-T}
\end{equation}
This reaction is modeled by a stochastic process that has formerly
been used in Brownian Dynamics simulations \cite{Ono:2001}. 
Inbetween every two steps of the numerical integrator for the DPD
equation of motion, each precursor dimer can be transformed into a
surfactant molecule with a spontaneous reaction rate $k_b$.
The spontaneous reaction can be catalytically enhanced by nearby
surfactants whose catalytic influence decreases linear with the 
distance to the reactant up to a certain threshold $r_{cat}$.
For simplicity, the effect of several catalysts is modelled as a
superposition:
\begin{equation}
    k = k_b + \sum_{i\in \mathbf C} k_s
\left\{
\begin{array}{cl}
    \left( 1 - \frac {r_{\mathbf C}} {r_{cat}}\right) & \mbox{ if } r<r_{cat}\\
    0 & \mbox{ otherwise}
\end{array}
\right.
\end{equation}
where $r_{\mathbf C}$ is the distance of the catalyst and $k_s$
the maximal catalytic rate per catalyst.
For clarity of the results, we set the spontaneous reaction rate in
our simulations to $0\tau^{-1}$.
In the upcoming simulations, $k_{cat}$ is set to $1.0\tau^{-1}$,
$r_{cat}$ to $1r_c$.
If a reaction occurs, the type of one random $\mathbf T$ bead is 
changed to $\mathbf H$, but positions and momenta are preserved. 

We also introduce particle exchange into the model to mimic the 
support of new precursors into the system: During the
simulation, two water beads can be exchanged by precursor dimers
with the probability $2.5\cdot10^{-5}\tau^{-1}$ per water bead
within a region of radius $2r_c$. Again, bead positions and momenta
are preserved.

\section{Results}

We implemented the DPD method using a velocity-Verlet integrator
($\lambda = 0.5$) with a step width of $0.04\tau$. The spatial domain 
is three dimensional, with periodic boundary conditions and 
size $(10r_c)^3$. In all the following simulations, the system is 
initialized with one surfactant dimer and 2998 water molecules 
adding up to a mean particle density of $3r_c^{-3}$. Unless 
otherwise specified, simulations have been run for 
$0\tau\le t\le1000\tau$ (25000 iterations).

\begin{figure}[tb]
        \centering
        \includegraphics[height=\columnwidth,angle=270]
                {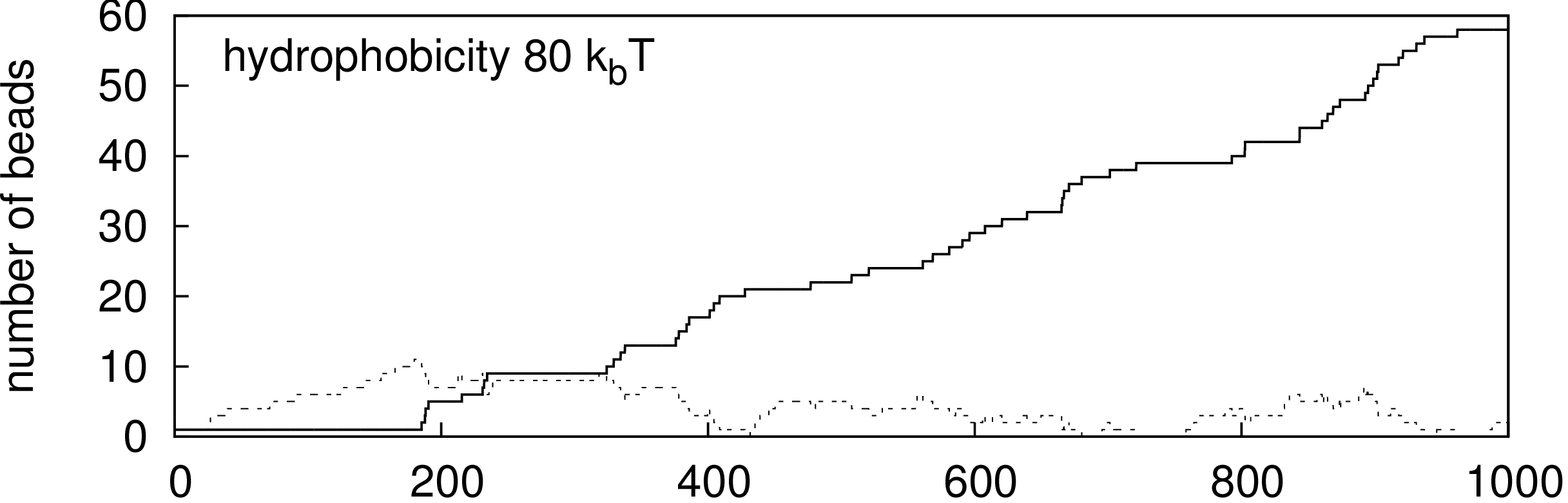}
        \includegraphics[height=\columnwidth,angle=270]
                {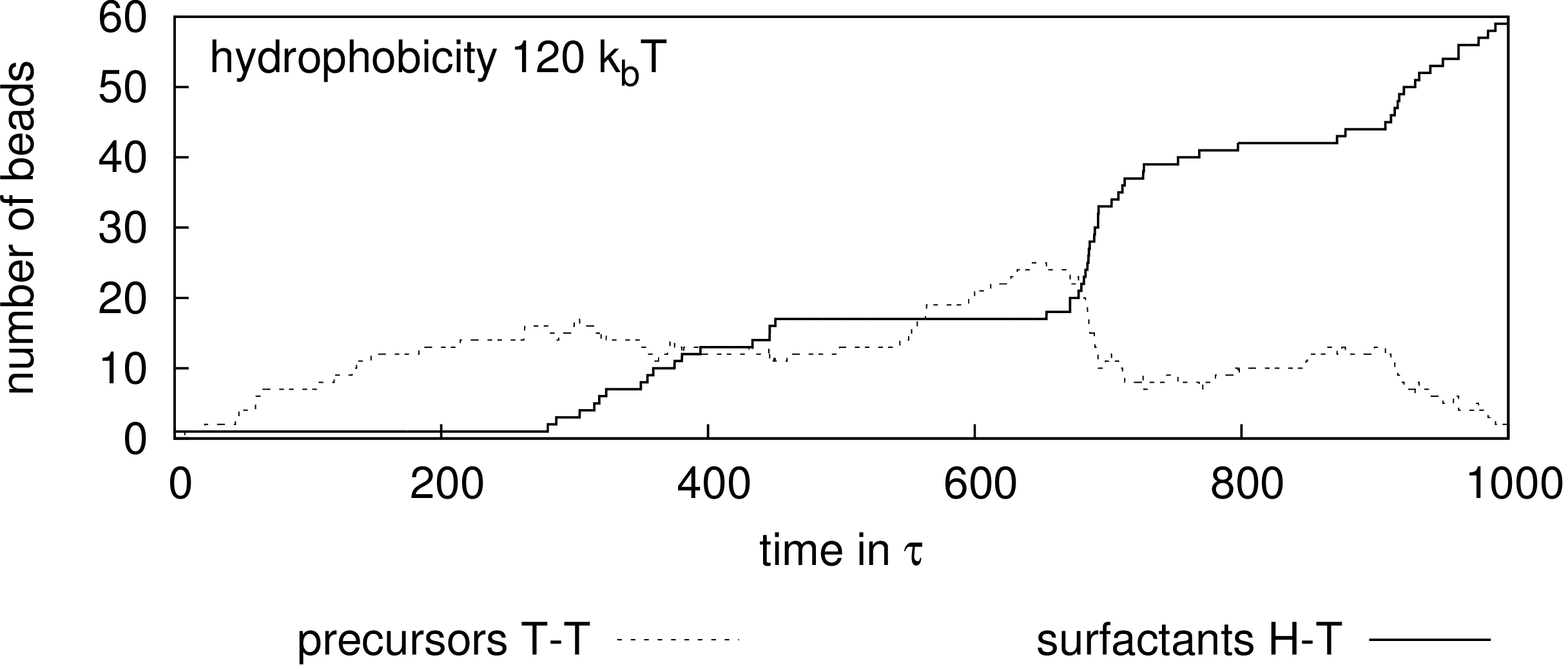}
        \caption[Evolution of bead numbers with time]{
                Evolution of bead numbers with time for two different
                hydrophobicity values ($a_{\mathbf H \mathbf T}$ and
                $a_{\mathbf W \mathbf T}$). The constant supply
                of precursors is counteracted by their transformation into
                surfactants. This transformation happens in spurts rather
                than continuously, as the precursor forms droplets in the
                aqueous solution. Size and frequency of these spurts depend
                on the hydrophobicity of the tail beads.
        }
        \label{fig_bead_evolution}
\end{figure}

For simulation runs with the parameter set given in the last
section, we typically get the following behavior: water beads are
successively exchanged by precursors in the exchange region of
the system. While diffusing through the system, precursors form
droplets due to their hydrophobic trait. Once such a droplet
reaches the initial surfactant, the precursors are turned
into surfactants. The turnover happens fast compared to supply
and diffusion of the precursor. New surfactants quickly arrange 
into a micellar shape with hydrophobic beads in the interior and 
hydrophilic beads towards the surface of the assembly. With
the parameters introduced before, this rearrangement takes only
few time steps and is thus fast compared to the metabolic
turnover. Such spontaneously formed nanocells diffuse through the 
system space as aggregates and eventually incorporate additional 
precursor droplets in their interior, where the metabolic process 
is repeated. The evolution of the overall system composition 
(number of beads per type) traces the different processes on their 
respective time scales (see figure \ref{fig_bead_evolution} for
$a_{\mathbf H \mathbf T}=a_{\mathbf W \mathbf T}=80k_bT$ (upper 
panel) and $a_{\mathbf H \mathbf T}=a_{\mathbf W \mathbf T}=120k_bT$
(lower panel). As one can see, the overall production of
surfactants is limited by a linear growth that results from the
constant supply of precursors. Locally, however, when a single
droplet is consumed by a nanocell, the metabolic turnover exhibits
a logistic growth which is suspected from resource limited
auto-catalysis. The logistic growth can be best seen in the lower
panel of figure \ref{fig_bead_evolution} around $t=700\tau$.

\begin{figure}[tb]
    \centering
    \includegraphics[height=\columnwidth,angle=270]{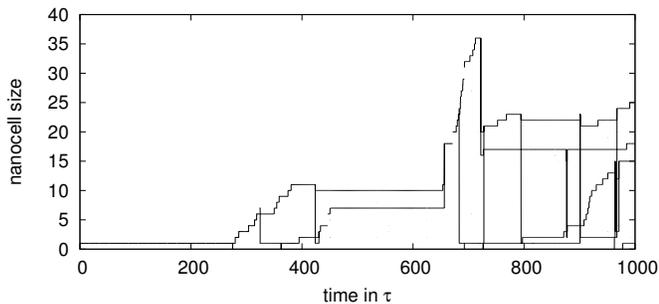}
    \caption[Size evolution with time]{
        Here, the size evolution of figure \ref{fig_bead_evolution}
        (lower panel) is tracked for individual nanocells: each line
        designates the size evolution of a single aggregate.
        Horizontal lines result from fission (or dissociation) events,
        after which two lines indicate the fate of the daughter cells.
        Isolated dots denote short term vibrations during which 
        nanocells divide and fuse within less than $1\tau$. For
        clarity, such horizontal lines have been suppressed for such
        vibrations. As one
        can see, only two daughter cells result from a true
        fission event (at $t=722\tau$). The other two result from
        dissociation of single surfactants that start to turn over
        precursor droplets found in bulk phase. These surfactant
        dissociations happen at $t=324\tau$ and $t=684\tau$. 
        Furthermore, two nanocells fuse at $t=655\tau$. 
    }
    \label{fig_micelle_sizes}
\end{figure}

\begin{figure*}[bt]
    \centering
    \includegraphics[width=.162\textwidth]{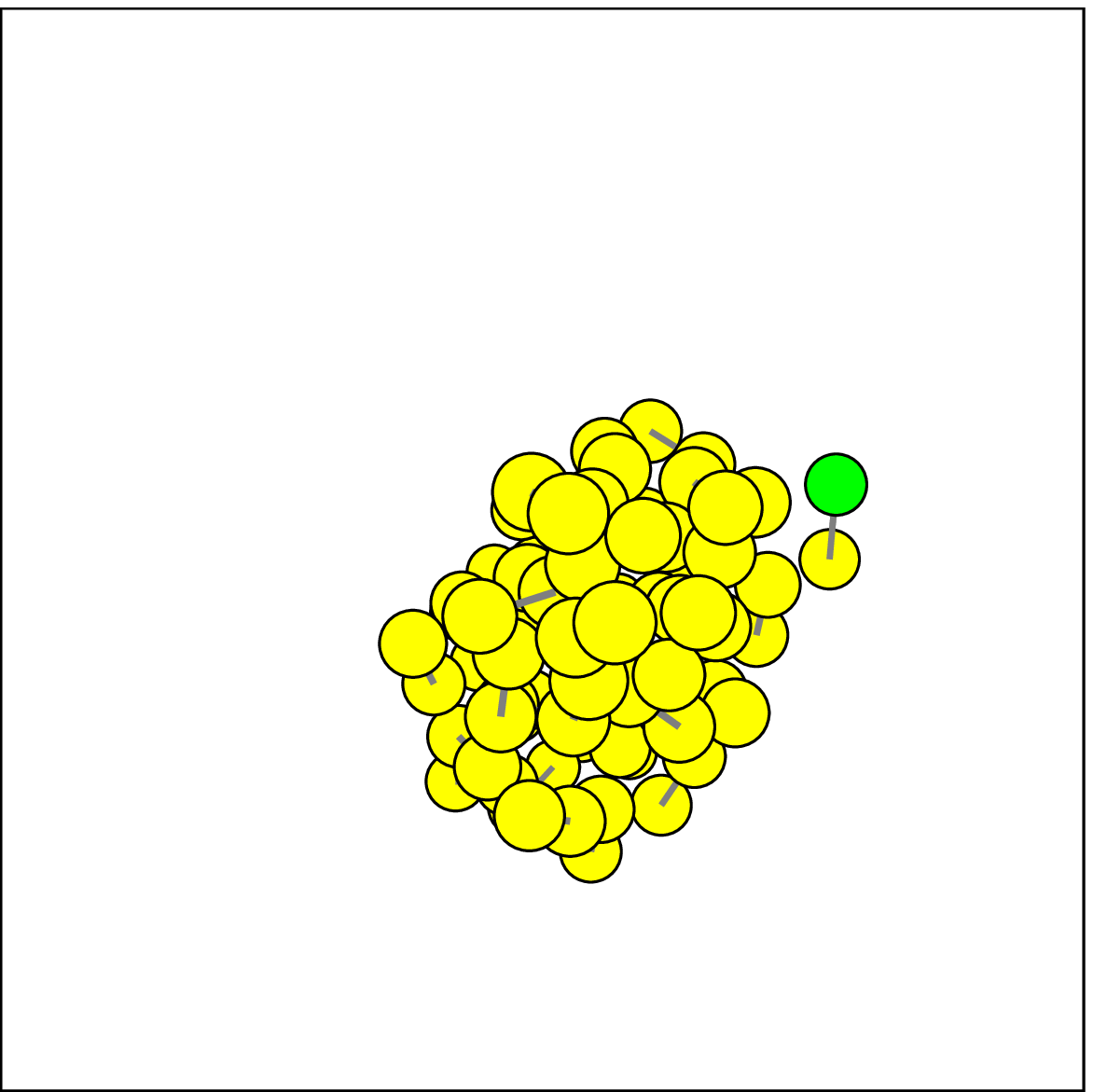}
    \includegraphics[width=.162\textwidth]{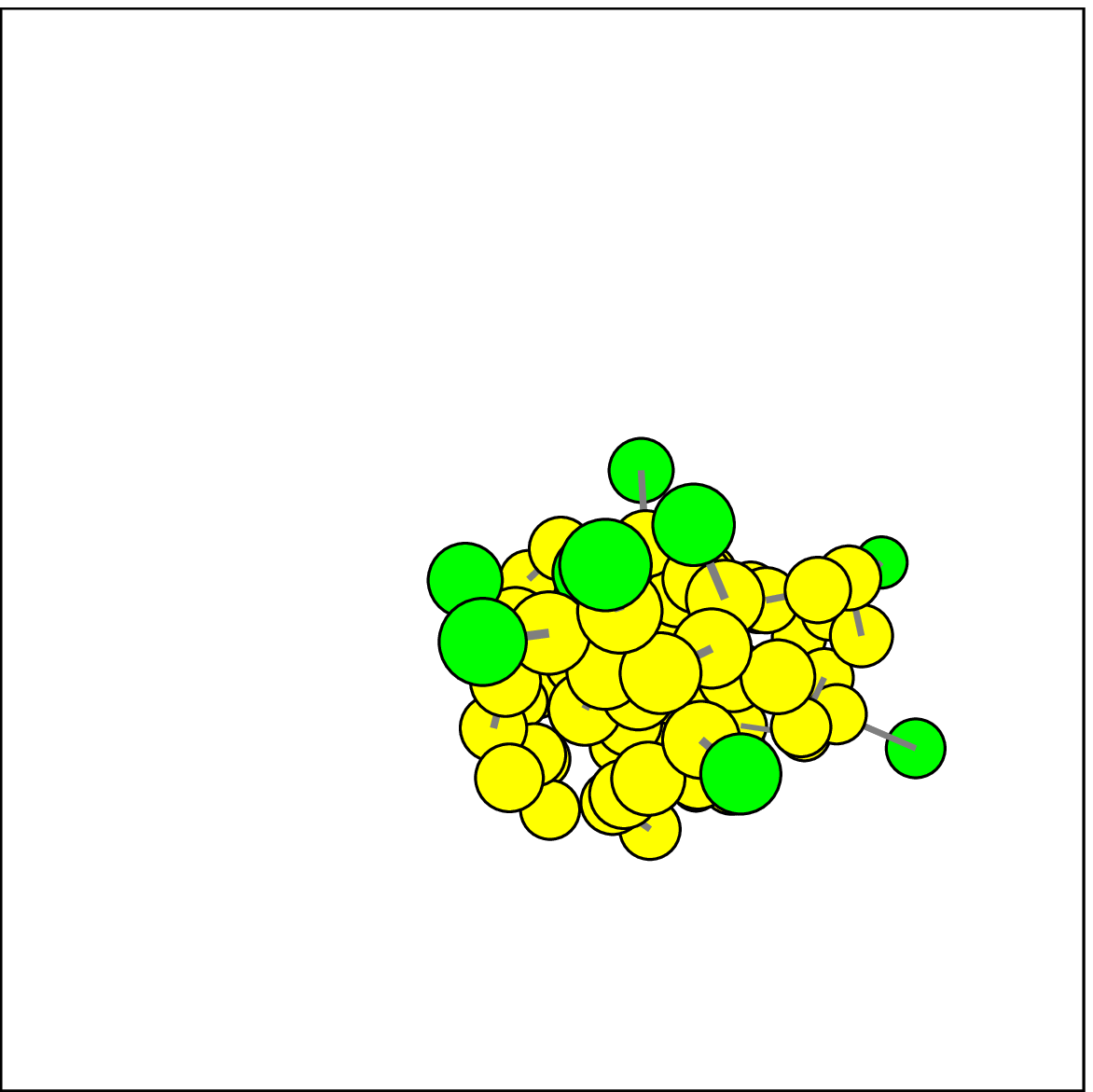}
    \includegraphics[width=.162\textwidth]{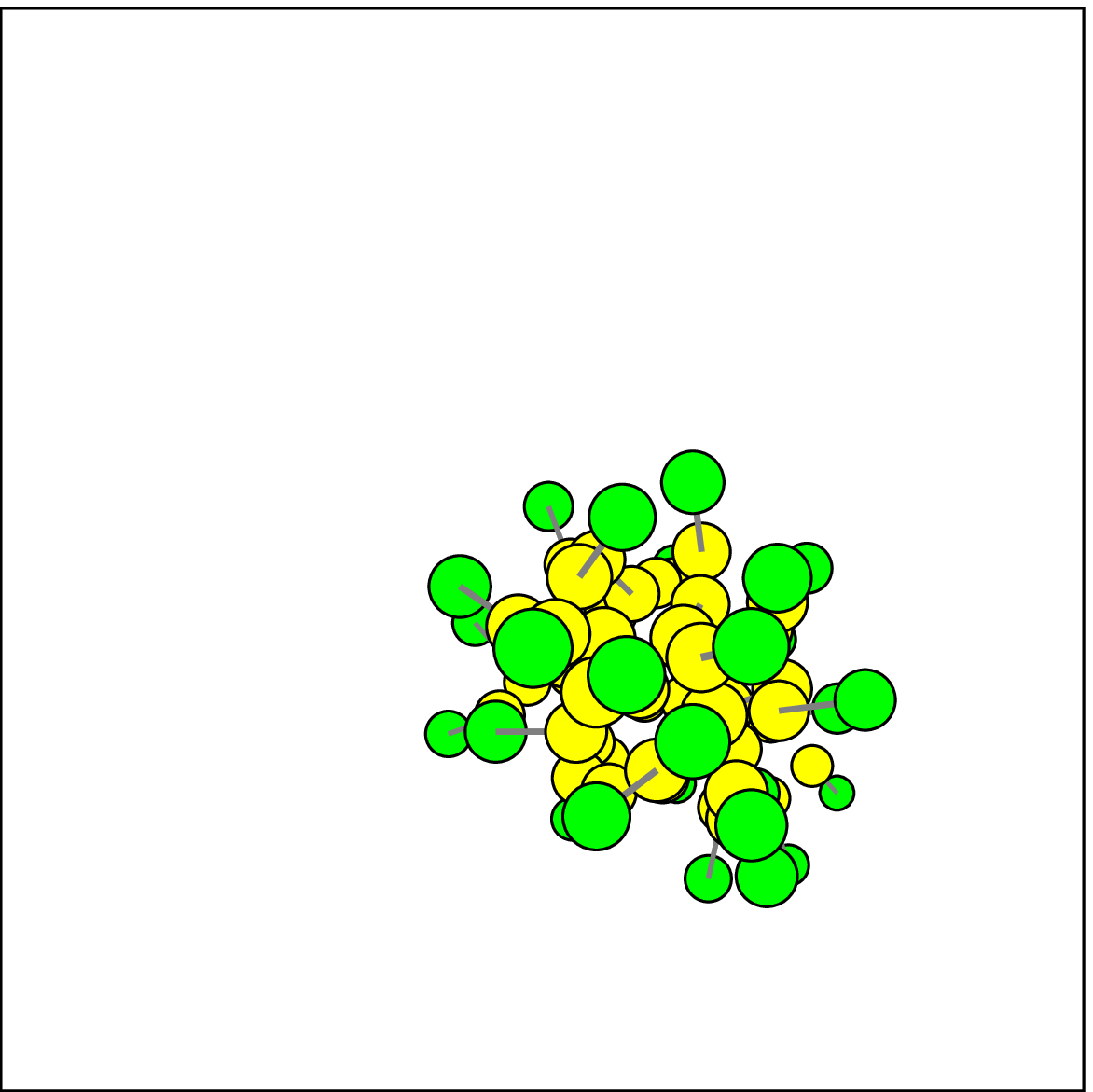}
    \includegraphics[width=.162\textwidth]{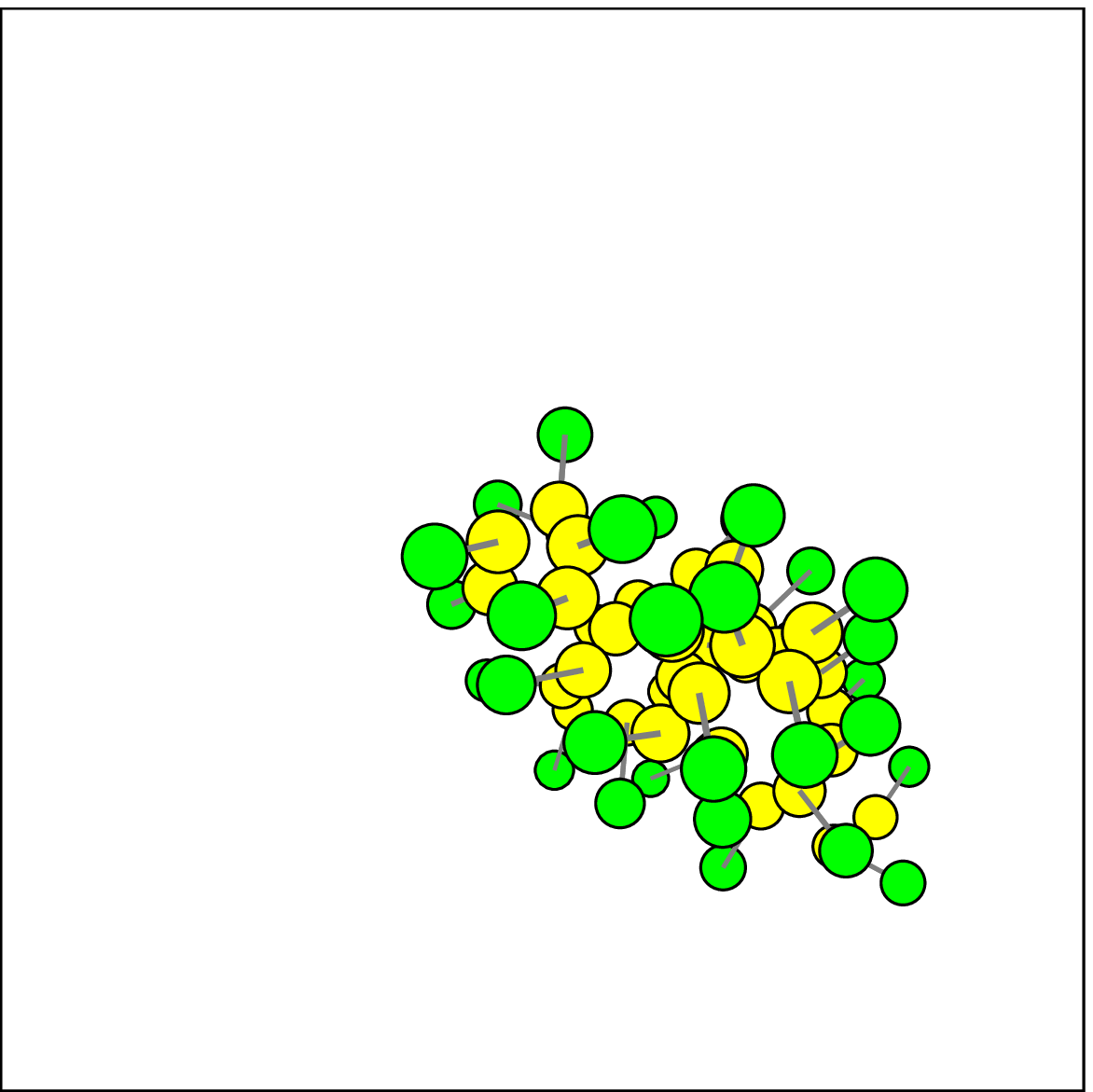}
    \includegraphics[width=.162\textwidth]{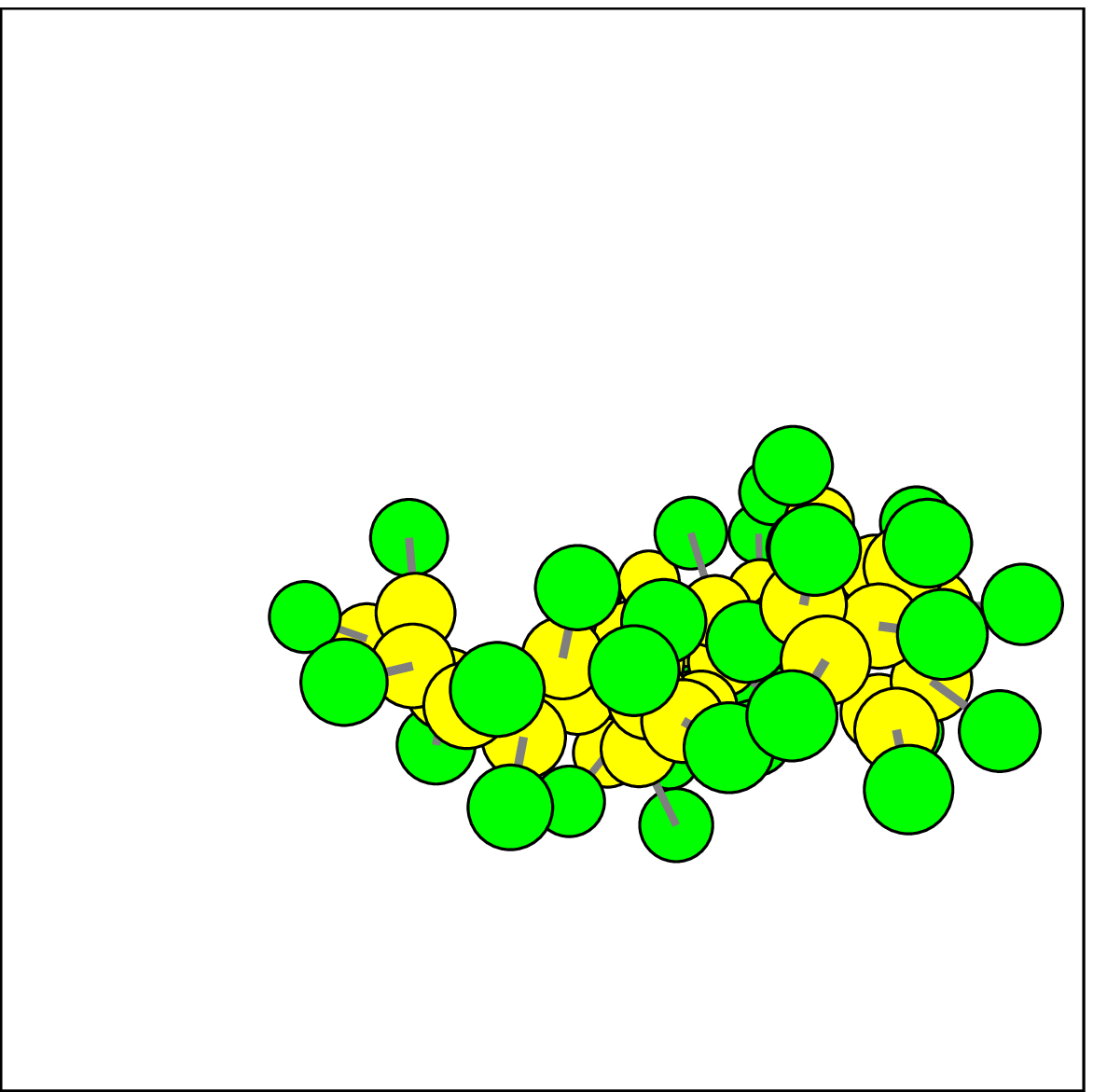}
    \includegraphics[width=.162\textwidth]{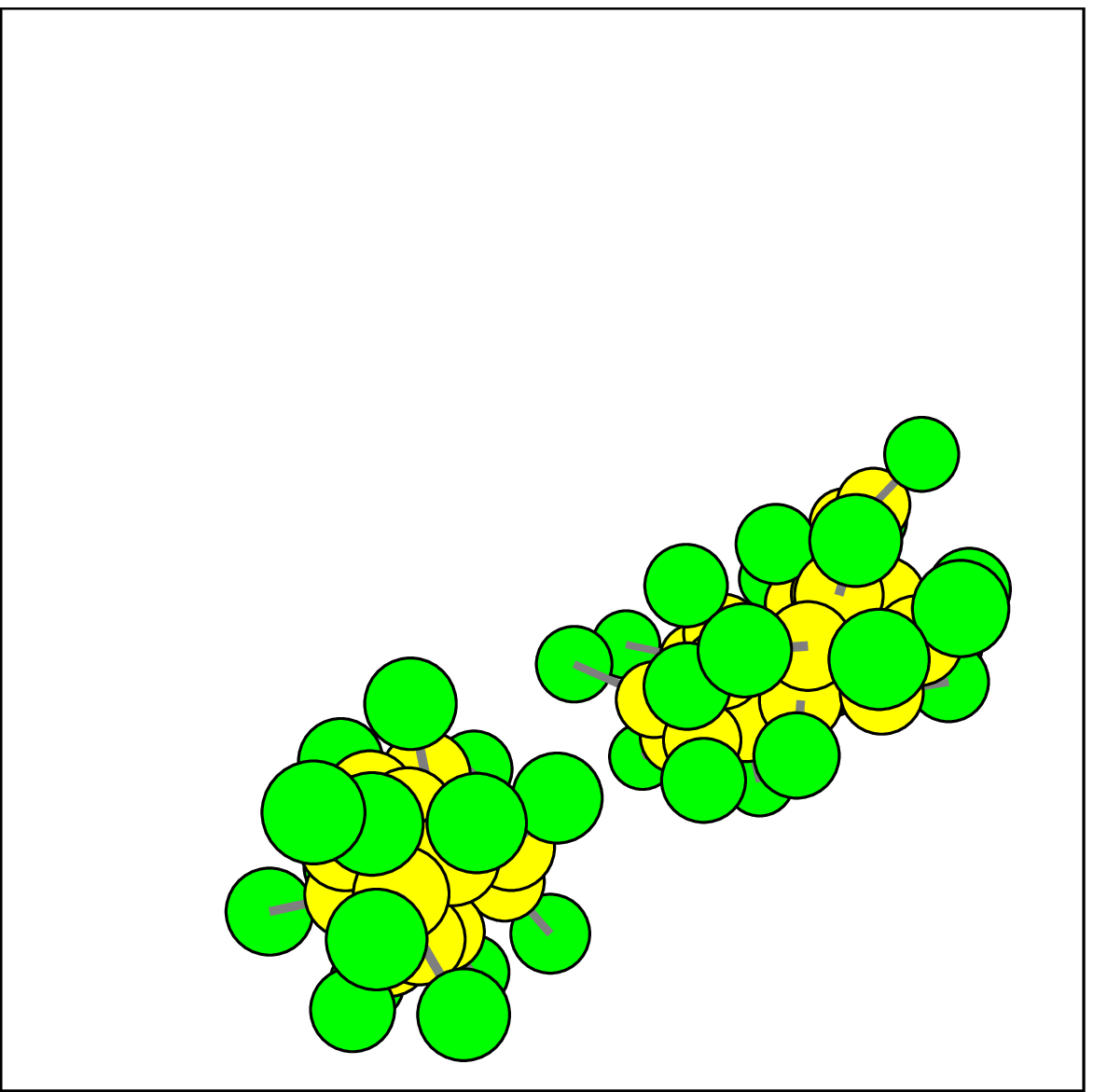}
    \caption[Metabolism and fission of a nanocell]{
        Metabolism and fission of a nanocell ($\mathbf T$ beads
        are shown in light, $\mathbf H$ beads in dark gray---water
        not shown):
        The initial surfactant metabolizes a precursor droplet and 
        turns it into a functioning nanocell (panels 1-3). While 
        the precursor is consumed, the nanocell elongates to 
        account for the changing precursor surfactant ratio (panel 
        4-5). Such elongated structures can be stable for several 
        time units, until---when all precursors are turned into
        surfactants---the nanocell divides into two daughter cells
        (last panel). 
    }
    \label{fig_fission}
\end{figure*}

On the level of individual micelles, the metabolic process 
increases the aggregate number, i.~e. the number of surfactants
per micelle. In a pure water-surfactant system, micelles would 
reject the surplus of surfactants into the bulk phase. In an 
oil-surfactant-water system, as the one under consideration, the 
hydrophobic core formed by the precursors, stabilizes the assembly
far beyond its original aggregate number. As a consequence, we
could observe that nanocells increase in aggregate number when
new surfactants are synthesized. While the precursor surfactant 
ratio shifts, the nanocell changes its shape from a spherical to
a rod-like micelle. When all or nearly all of the precursor is 
turned into surfactant, the nanocell finally becomes unstable and 
divides into two smaller aggregates (see figures 
\ref{fig_micelle_sizes} and \ref{fig_fission}). The nanocell 
division occurs in the cylindrical middle-part of the rod-like 
aggregate by indentation of surfactant heads. It induces vibrating 
modes into the daughter aggregates as they rearrange back to a 
spherical shape. Although this vibration is successively dissipated 
into undirected motion, it sometimes leads back to short series of 
temporary fusion and fission of the daughter cells.


Elongated micellar structures are well-known from worm-like
micelles which usually consist of two surfactants with different
curvature. Such worm-like micelles are stable equilibrium aggregates
and exhibit an exponential size distribution \cite{Kro:1996}. In
the system studied in this work, however, the elongated aggregate is
not stabilized by different curvature values of the components, but
by the hydrophobic core. Accordingly, once the precursor is turned
over into new surfactant molecules, the elongated structure looses
its stability.

There is a second pathway, however, that might jeopardize the above
scenario: once in a while throughout our simulations, nanocells 
loose individual surfactants into the bulk phase. If this relaxation
process happens fast compared to the metabolic turnover, the 
nanocells might not be able to reach the division size. Surfactants 
in the bulk phase may however metabolize precursor droplets and
spontaneously form nanocells on their own. 

Formally, fission events can be written in the form of a chemical 
reaction:
\begin{equation}
        S_{n+m} \stackrel {k^-_{m,n}} \longrightarrow S_n + S_m
        \label{eqn_fission}
\end{equation}
where $S_n$, $S_m$ and $S_{n+m}$ are aggregates of size $n$, $m$,
and $n+m$, respectively, and $k^-_{m,n}$ is the fission rate. 
For $m=1$, one obtains dissociations as a special case. Analogously, 
association and fusion events take the form
\begin{equation}
        S_n + S_m \stackrel {k^+_{m,n}} \longrightarrow S_{n+m}
        \label{eqn_fusion}
\end{equation}
In order to quantify micellar fission and surfactant dissociations, 
nanocells have been identified by a variant of the flood fill 
algorithm: every two $\mathbf T$ beads within a distance of $1r_c$ 
or less have been considered to belong to the same aggregate. 
The aggregate number is defined as the number of participating 
surfactants. This allows for tracking individual aggregate 
sizes and their transitions through time. Each of the above
reaction schemes results in two transitions, given by
\begin{equation}
        S_n \rightarrow S_{n+m} \quad;\quad S_m \rightarrow S_{n+m}
    \label{eqn_fission_transitions}
\end{equation}
and
\begin{equation}
        S_{n+m} \rightarrow S_n \quad;\quad S_{n+m} \rightarrow S_m
    \label{eqn_fusion_transitions}
\end{equation}
Furthermore, the turnover of precursors results in the transition
\begin{equation}
        S_n \rightarrow S_{n+1}
\end{equation}

Not all transitions can be expressed by the chemical reaction
scheme given by \ref{eqn_fission} and \ref{eqn_fusion}. Those
transitions are of the type
\[
        S_{n+m} + S_l \longrightarrow S_n + S_{m+l}
\]
as well as fissions into and fusions from more than two aggregates.
For all the simulations performed, such outliers amount to less than 
0.3\% of the total transitions. They have been neglected for 
further analysis.

In the remainder of this work, we will analyze two key parameters
of the model and their influence on the dynamics of the system. 
Thereby, we will put our attention on the fission pathways discussed
above.
The first parameter we analyze is the hydrophobicity, i.~e. the
values $a_{\mathbf T\mathbf W}$ and $a_{\mathbf T\mathbf H}$.
This parameter is related to the dissimilarity between surfactant
tails and water. For most surfactants the hydrophobicity is solely 
a function of the length of the hydrocarbon chains \cite{Eva:1999}.
The value is thus easily adjustable in experimental setups.

Second, we analyze the influence of the catalytic rate $k_c$--- 
hence, the speed of metabolic turnover---on the division pathway of 
the nanocell. As we did not specify the molecular implementation of 
precursor and surfactant, it is conceivable that in an experimental 
setup, one can choose from a wide range of catalytic efficiencies.

\subsection{Influence of hydrophobicity on nanocell dynamics}

\begin{figure}[tb]
        \centering
        \includegraphics[width=\columnwidth]
                {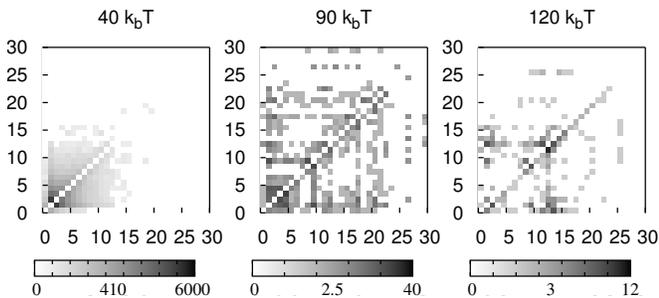}
        \caption[Histograms of nanocell size transitions]{
                Histograms of nanocell size transitions for different
                hydrophobicities (see text on how transitions are defined).
                For a transition $S_n\rightarrow S_m$, $n$ is ordered along
                the vertical, $m$ along the horizontal axis. Colors indicate
                the number of occurences. Note that colors have been scaled 
                by root functions to emphasize seldom events. 
        }
        \label{fig_transition_histograms}
\end{figure}

For three hydrophobicities, $40$, $90$ and  $120k_bT$, 
histograms of such transitions are shown in figure 
\ref{fig_transition_histograms}. The figures reveal a clear trend
both in aggregate numbers as well as transition types.
For hydrophobicity $40k_bT$, the system is almost entirely composed
of single surfactants and small aggregates in bulk phase. 
33.4 \% of the transitions are dissociations and associations of 
two single surfactants. For the few bigger aggregates, transitions 
are distributed more or less homogeneously, i.~e. surfactant 
dissociation is as likely as proper aggregate fission. Thus, for 
weak hydrophobicities, the system resembles a homogeneous solution 
without significant formation of structures.
For $120k_bT$, on the other hand, the transition histogram looks
completely different. Associations or dissociations of two isolated
surfactants make only 3.7\% of the transitions, for this parameter.
The most prominent transition type is the turnover of a precursors
within nanocells that range in size from 1 to 25 
surfactants (20.6\% of all transitions), represented by high values 
in the lower secondary diagonal.
The absence of an upper secondary diagonal reveals that there are 
no surfactant dissociations except some between aggregate numbers 
8 and 16. Due to the higher stability of aggregates, there are far 
less overall transitions than in the previous case (321 compared
to 24743), which poses a problem when trying to obtain smooth 
histograms. Although there are distinct cases of proper fission
events (e.~g.  $S_{25}\rightarrow S_{14}+S_{11}$, 
$S_{20}\rightarrow S_{11}+S_9$) it is hard to tell from the
histograms whether such fission is more likely than single 
surfactant dissociation. Varying the hydrophobicity between these 
two extrema yields traits of both other histograms. One example is
given in figure \ref{fig_transition_histograms} for $90k_bT$.

\begin{figure}[tb]
        \centering
        \includegraphics[width=\columnwidth]
                {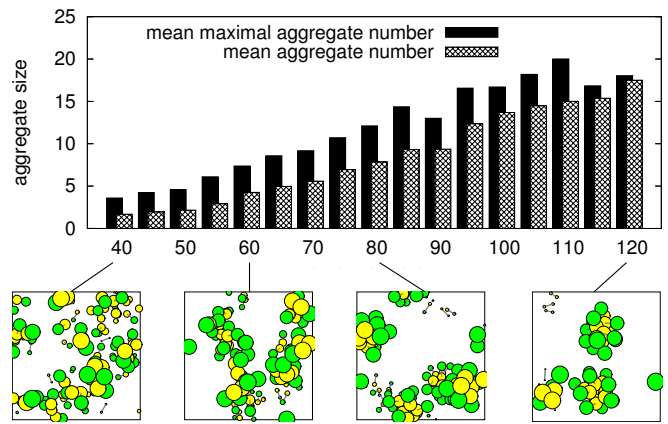}
        \caption[Mean aggregate numbers for different hydrophobicity values]{
                Mean aggregate and mean maximal aggregate numbers 
                for hydrophobicity values between $40$ and $120k_bT$.
                The latter averages only the size of nanocells which
                are actually going to divide, while the former averages
                all aggregates. System states have been averaged for
                $500\tau\le t\le 1000\tau$. Earlier states have been considered
                as transient. Below, parts of the final simulation states 
                are shown for selected parameters.
        }
        \label{fig_ht_vs_meansize}
\end{figure}

As new precursors are constantly supplied and nanocells grow and
divide over time, it is somewhat difficult to capture mean
aggregate numbers of the assemblies. Nevertheless, these values
are prominent characteristics in the study of micellar systems and 
their knowledge can help to get insight into the system under 
consideration. To gain aggregate numbers we compute
the average value $m+n$ for all transitions of the form
$S_{m+n}\rightarrow S_n+S_m$, i.~e. we average aggregate numbers
only in moments of fission or dissociation. The results can be
regarded as the mean maximal aggregation number of nanocells.
Figure \ref{fig_ht_vs_meansize} shows results for hydrophobicity 
values from $40k_bT$ to $120k_bT$ together with a simple average 
of all aggregate sizes in the system.
Both maximal and average values increase constantly from 1.68
(3.59) for $40k_bT$ to 18.04 (17.51) for $120k_bT$. It becomes
apparent, that for very weak hydrophobicities most of the 
surfactants are either isolated in bulk phase or in very small 
assemblies. For strong hydrophobicities, aggregates are very
distinct and single surfactants in bulk phase are rare.
There is, however, no sharp boundary or phase transition between
small sub-micellar assemblies and proper micelles, but rather a 
continuous transition.

\begin{figure}[tb]
        \centering
        \includegraphics[height=\columnwidth,angle=270]
                {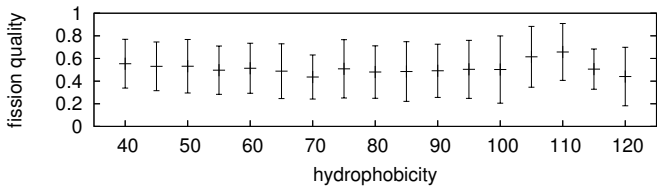}
        \caption[Mean quality of nanocell divisions]{
        Mean quality of nanocell divisions. A value of 1 signifies
        division into daughter cells of equal size, while a value
        close to 0 results from single surfactant dissociation.
        Error bars denote standard deviations.
        }
        \label{fig_fission_quality}
\end{figure}

Going back to the issue of nanocell division, we want to
distinguish proper fission into nanocells of approximately equal
size from dissociation of sub-micellar aggregates. The previous 
analysis revealed that one cannot use absolute aggregate numbers, 
as they exhibit a trend for stronger hydrophobicity. Therefore, we 
characterized each event of the form of equation \ref{eqn_fission} 
by the function
\begin{equation}
    Q(m,n) = 1 - \frac {\left|m-n\right|} {m+n}
\end{equation}
which denotes the relative fission quality. $Q(m,n)=1$ for $m=n$, 
i.~e. when the resulting nanocells are equal in size. $Q$ becomes
smaller as daughter cells become less alike. We have averaged $Q$ 
as a function of the hydrophobicity over all fission events in the 
simulation runs discussed before. Mean and standard deviation are 
shown in figure \ref{fig_fission_quality}.
$\left<Q\right>$ varies between $0.41$ and $0.65$ with no 
significant trend for weak or strong hydrophobicities. Moreover,
standard deviation is very high. This reveals that fission into
any two daughter cells is equally probable, no matter the ratio
of their sizes. In terms of fission rates this finding can be
written as
\begin{equation}
    k^-_{m,n} =: k^-(m+n)
\end{equation}
for the system under consideration.

\subsection{Influence of the catalytic rate on nanocell dynamics}

\begin{figure}[bt]
    \includegraphics[height=\columnwidth,angle=270]
        {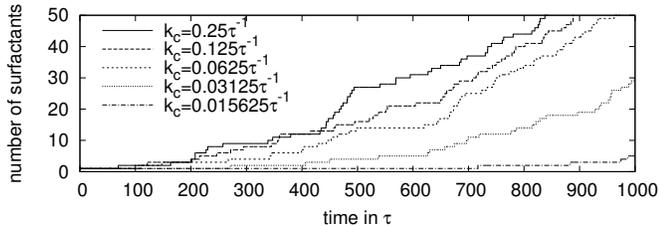}
    \caption[Number of surfactants as a function of time]{
        Number of surfactants as a function of time for different
        catalytic rates $k_c$. For slow metabolic turnover, the
        exponential shape of the auto-catalysis becomes apparent.
    }
    \label{fig_kc_vs_surfactants}
\end{figure}

The catalytic rate has been varied from $2^{-6}\tau^{-1}$ to
$4.0\tau^{-1}$ in exponential steps. Hydrophobicity has been set
to $80k_bT$. Global surfactant dynamics are shown in figure 
\ref{fig_kc_vs_surfactants}.
For slow metabolic turnover ($<0.125\tau^{-1}$), the exponential 
shape of surfactant production becomes apparent, i.~e. the 
constant precursor supply of precursors does not limit surfactant
production over the simulated time span. This is tantamount to 
saying that unmetabolized precursor droplets are present throughout 
the whole simulation. Deceleration of the global dynamics is 
reflected in slower fission rates of individual nanocells. For
example, we observed that the first fission event is retarded by 
$180\tau$ on the average when catalytic rates are halved. For 
$k_c < 0.03125\tau^{-1}$, no fission occurs during the simulated 
time span.

\begin{figure}[bt]
    \includegraphics[height=\columnwidth,angle=270]
        {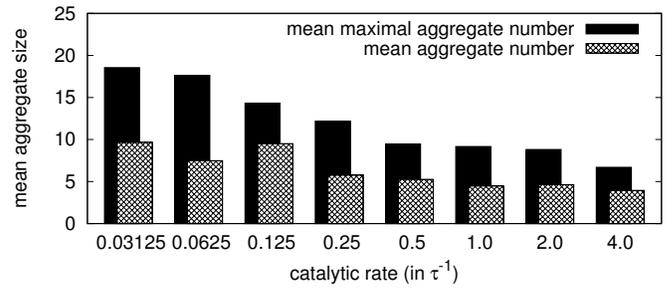}
    \caption[Mean aggregation numbers for different catalytic rates]{
                Mean aggregate and mean maximal aggregate numbers 
                for catalytic rates between $0.03125$ and $4.0\tau^{-1}$.
        Consult the caption of figure \ref{fig_ht_vs_meansize} on
        how aggregation numbers have been obtained.
    }
    \label{fig_kc_vs_meansize}
\end{figure}

Figure \ref{fig_kc_vs_meansize} shows mean aggregation numbers as 
a function of the catalytic rate.
As one can see, slow metabolic turnover increases both maximal and
average aggregation numbers (from 9.45 (5.27) for $0.5\tau^{-1}$ to 
18.55 (9.66) for $0.03125\tau^{-1}$). The trend becomes less
apparent for faster metabolic turnover ($1.0\tau^{-1}$ to 
$4.0\tau^{.1}$). For $k_c=0.125\tau^{-1}$, the maximal aggregation 
number is slightly higher than a monotonic trend would imply---a 
fact that we relate to statistical deviations, as fission events are 
considerably rare for small catalytic rates. The increase in both
aggregation numbers is a natural consequence of the decelerated
metabolism: when the precursor is metabolized slowly while its 
supply is held constant, the size of the hydrophobic core increases,
and offers a bigger area for surfactants to attach. Hence, the 
maximal aggregate numbers increase.

\begin{figure}[bt]
    \includegraphics[height=\columnwidth,angle=270]
        {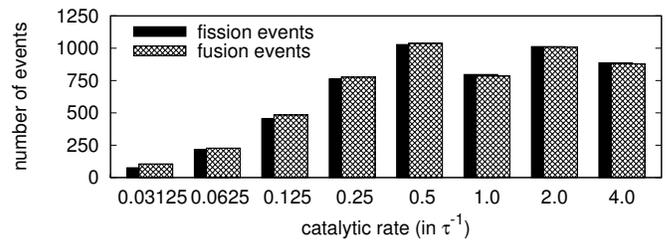}
    \caption[Number of overall fission and fusion events]{
        Number of overall fission and fusion events 
        (transitions of the form of equations
        \ref{eqn_fission_transitions} and 
        \ref{eqn_fusion_transitions}) as a function of the 
        catalytic rate $k_c$. 
    }
    \label{fig_kc_vs_dissociations}
\end{figure}

For the above runs, the number of fission and fusion events has
been measured (see figure \ref{fig_kc_vs_dissociations}).
For all simulation runs, fusion and fission events are more or less
balanced. This reveals that most of these events result from 
surfactant exchange with the bulk phase or from series of temporary 
fission and fusion during a single division process rather than 
from proper nanocell divisions. For low catalytic rates 
($0.03125\tau^{-1} \le k_c \le 0.25\tau^{-1}$) the number of such 
balanced transitions falls significantly from 1028 for 
$k_c=0.5\tau^{-1}$ to only 8 for $k_c=0.03125\tau^{-1}$. As in the 
case of aggregation numbers, the trend in transition numbers can be 
related to the hydrophobic core: the more hydrophobic particles in 
the interior of a nanocell, the less dissociations occur on its 
surface. A strong anti-correlation between aggregation numbers and 
the number of fission/fusion events (with a correlation coefficient 
of $-0.917$) justifies this hypothesis.

It has to be pointed out, however, that the catalytic rate might 
affect nanocellular dynamics only during a certain transient time. 
It has been shown how the nanocellular dynamics depend on the ratio 
between metabolic turnover and precursor supply. Precursors are 
supplied by diffusion. Therefore, the rate of their incorporation 
into an individual cell depends on the overall concentration of 
nanocells. Once a critical cell concentration is reached, precursor 
incorporation might be slower than its metabolic turnover, which 
would undermine the above discussed effect. While the duration of
such transient will depend on the ratio of precursor supply and
turnover, dynamics after the transient might be little affected 
again. Ideally, simulations would be performed in a homeostatic system, 
with an influx of precursor solution at one side and an outflux
of reaction products at the opposite site of the system. Up to now,
however, little is known about the performance of DPD in such open
systems.

\section{Discussion}

In this paper we have presented an information-free nanocell based
on a micellar system and a single auto-catalytic reaction that 
serves as metabolism. This simple system can be understood as a 
minimal self-replicating chemical system. As such, it denotes the 
boundary between pure auto-catalysis and a more complicated 
self-reproducing system which would also include inheritable 
information. We analyzed the dynamics of this nanocell using a 
dissipative particle dynamics approach. This simulation technique can cover 
the relevant time scale, while it has been shown to be still 
physically accurate compared to other simulation techniques. As a 
consequence, we have been able to perform analyses of the system
in a level of detail, that has---as far as we know---not been
reached before in the study of self-replicating entities.

The general replication cycle of micellar nanocells by metabolic
turnover and division is very robust against changes in
hydrophobicity and catalytic rates. 
It has been shown that the mean aggregation number of nanocells 
depends on the hydrophobicity of the surfactant (and precursor)
as well as on the catalytic rate of the metabolism. For increasing 
hydrophobicity, a monotonic change in aggregation number with no 
sudden phase transition has been observed, ranging from a 
nearly homogeneous solution with only submicellar aggregates for 
weak hydrophobicities to the formation of distinct micelles in 
surfactant-free water for a very high hydrophobicity. The same 
monotonic increase in aggregation number could be observed for
increasing catalytic rates, i.~e. fast metabolic turnover.

It has been found that the rate of nanocell fission and surfactant
dissociation depends on the size of the hydrophobic core of the 
nanocells, and is more likely to occur for small values in hydrophobicity 
and slow metabolic turnover. Daughter cells resulting 
from a fission event have been shown to vary significantly in 
size. There is neither a trend in the average size ratio of fission 
products nor in its variance. 

Our work shows that the envisioned replication-cycle 
of nanocells---namely incorporation and turnover of precursor 
droplets followed by eventual aggregate division---is achievable 
over a wide range of parameters. In fact, there is 
no parameter combination for which the general 
replication cycle has been rendered impossible: although we have
been able to decrease mean aggregation numbers and increase
dissociation rates up to a point where the system obeys no clear
structures, we could not observe that dissociation of single 
surfactants jeopardizes the growth of a otherwise stable nanocell.

We have demonstrated the feasibility of a self-replicating system
in the absence of genetic information. Although such a system is
not able to evolve towards higher complexity, it could have served
as a functioning subsystem providing metabolism and embodiment for 
subsequent protocells of higher evolutionary complexity. It is
conceivable, that independently evolved information systems like 
RNA might have become incorporated into such functioning
replicators. When the two formerly independent replication cycles 
of container and genome are orchestrated by coupling, such that
each daughter cell of the dividing container is loaded with exactly
one copy of the genomic information, one would obtain a true
self-reproducing protocell with the ability to metabolize, divide 
and evolve. 

Apart from prebiotic scenarios in which micelles are considered
as possible ancestors of subsequent vesicle based organisms, such 
systems are explored in the context of so-called living technology, 
i.~e. artificial systems that mimic life-like behavior. 
Our results suggest that the generic replication cycle of micellar 
nanocells is a robust basis for artificial life forms.
We are currently exploring one design of such an artificial 
protocell in which genomic information is coupled to a micellar 
self-replicating system similar to the one presented here.

\vspace{0.5 cm}

{\bf Acknowledgments}

\vspace{0.2 cm}
The authors would like to thank the members of the Complex Systems 
Lab for useful discussions. This work has been supported by grants 
FIS2004-0542, IST-FET PACE project of the European Community 
founded under EU R\&D contract FP6002035 and by the Santa Fe Institute.
\vspace{0.2 cm}

\bibliography{references}

\end{document}